\documentclass[preprint,12pt]{elsarticle}

\usepackage{amsmath}
\usepackage{amssymb}
\usepackage{amsfonts}
\usepackage{graphics}
\usepackage{graphicx}
\usepackage{amscd}
\usepackage{amsfonts}
\usepackage{epsf}
\usepackage{epsfig}
\usepackage{color}

\def\m#1{\mathrm{#1}} 
 
\def\tendto{\rightarrow} 

\def\cbra#1{ \langle {#1}|}
\def\cket#1{|{#1}\rangle}
\def\be#1{\begin{equation}#1\end{equation}} 

\def\integral#1#2{\int^{#2}_{#1}}

\begin{document}

\begin{frontmatter}

\title{Lattice study of supersymmetry breaking in ${\cal N}=2$ supersymmetric quantum mechanics}

\author{Daisuke Kadoh}
\address{Department of Physics, Faculty of Science, Chulalongkorn University, Bangkok 10330, Thailand\\
Research and Educational Center for Natural Sciences, Keio University, Yokohama 223-8521, Japan}
\ead{kadoh@keio.jp}

\author{Katsumasa Nakayama}
\address{Department of Physics, Nagoya University, Nagoya, 464-8602, Japan\\
KEK Theory Center, High Energy Accelerator Research Organization (KEK), Tsukuba 305-0801, Japan
}
\ead{katumasa@post.kek.jp}

\begin{abstract}
We study supersymmetry breaking from a lattice model of ${\cal N}=2$ supersymmetric quantum mechanics 
using the direct computational method proposed in arXiv:1803.07960. 
The vanishing Witten index is realized as a numerical result in high precision. 
The expectation value of Hamiltonian is evaluated 
for the double-well potential. Compared with the previous Monte-Carlo results,
the obtained vacuum energy coincides with the known values
within small errors for strong couplings.  The instanton effect is also reproduced for weak couplings. 
The used computational method helps us to evaluate the effect of finite lattice spacings more precisely 
and to study the mechanism of non-perturbative supersymmetry 
breaking from lattice computations.
%The methodology for defining Hamiltonian on the lattice is also examined.
\end{abstract}


\end{frontmatter}

\section{Introduction}
\label{sec:introduction}
Supersymmetry (SUSY) plays an important role in physics beyond the standard 
model and a quantum theory of gravity such as superstring theory, 
while it should be broken at a low energy scale.
So various mechanisms that trigger SUSY breaking have been studied 
with great interest for a long time \cite{Witten:1981nf,Witten:1982df}.
Since SUSY can be spontaneously broken by a non-perturbative effect, 
the lattice computation is expected to be useful 
to study the breaking mechanism in detail \cite{Beccaria:2004pa,Kanamori:2007yx, Wozar:2011gu, Catterall:2015tta, Catterall:2017xox}.

The lattice approach to SUSY, however, suffers from several 
problems stemmed from two fundamental issues, 
no infinitesimal translation on the lattice and the sign problem of the Monte-Carlo method.
The absence of the infinitesimal translation is directly 
associated with the explicit SUSY breaking 
of the lattice action and no explicit definition of Hamiltonian, 
while the sign problem is a practical issue of the Monte-Carlo 
method, from which general SUSY theories suffer.

In order to control the artificial SUSY breaking due to the lattice cut-off, various studies have been carried out so far
\cite{ Dondi:1976tx,
Elitzur:1982vh,
Elitzur:1983nj,
Cecotti:1982ad, 
Sakai:1983dg,
Golterman:1988ta, 
Catterall:2000rv,
Kikukawa:2002as,
Giedt:2004qs,
Feo:2004kx,
Kato:2008sp,
Kadoh:2009sp,
Kadoh:2010ca,
Kato:2013sba,
Kadoh:2015zza, 
Asaka:2016cxm,
Kato:2016fpg,
DAdda:2017bzo,
Kadoh:2018ele}. 
In ${\cal N}=2$ supersymmetric quantum mechanics (SUSY QM), lattice models with 
non-local SLAC derivative and improved Wilson term and one-exact supersymmetric 
action have been studied\cite{ Catterall:2000rv,
Giedt:2004qs,
Bergner:2007pu,
Schierenberg:2012pb}.
The similar lattice models were also proposed in two-dimensional Wess-Zumino model
\cite{ Beccaria:1998vi,
Catterall:2000rv,
Catterall:2001fr,
Giedt:2004vb,
Giedt:2005ae,
Kanamori:2007yx,
Bergner:2007pu,
Kastner:2008zc,
Bergner:2009vg,
Kawai:2010yj, 
Kanamori:2010gw,
Kamata:2011fr,
Wozar:2011gu,
Schierenberg:2012pb,
Steinhauer:2014yaa}
\footnote{For several attempts in gauge theories, see the references of \cite{Kadoh:2016eju}.}.
With a few exact supersymmetries, an appropriate lattice  Hamiltonian 
can be defined such that the zero point energy vanishes \cite{Kanamori:2007yx}. 
% thanks to the exact supersymmetry.
Thus, one can say that the first issue which is related to the translational 
invariance is resolved at least for low dimensional models. 

On the other hand, it is not straightforward to apply the Monte-Carlo method to models with supersymmetry breaking.
For the periodic boundary condition, the partition function (Witten index) vanishes, 
and the expectation values are formally ill-defined without any regularization.
Even if we impose the anti-periodic boundary condition on the fermions to avoid this issue, 
it is quite difficult to study the SUSY breaking triggered by an instanton effect 
to which the Monte-Carlo method is not readily applicable.
Even for SUSY QM, the importance sampling is hard 
for the double-well potential unless very strong couplings
are taken since SUSY is broken by rare tunnelings 
between two minima of the double-well \cite{Kanamori:2007yx,Kanamori:2010gw}.

Recently the authors have proposed an alternative computational method based 
on the transfer matrix in SUSY QM \cite{Kadoh:2018ele}, which is closely 
associated with the tensor network approach applicable to higher dimensional field theories. 
The correlation functions can be computed as a product of transfer matrices, and any stochastic process is not needed. 
Thus the numerical results are given without statistical uncertainty, and the sign problem never exists in the first place. 
For theories without SUSY breaking, we have found that numerical results reproduce the known results. 
However, with SUSY breaking, it is still unknown whether this method is useful or not.
In order to investigate SUSY breaking in higher dimensional models, further numerical studies 
in SUSY QM with this method are needed.

This paper presents a lattice computation of the vacuum energy for the double 
well potential ($\phi^4$ theory) with SUSY breaking using the transfer matrix method.
We employ the lattice action with one exact supersymmetry \cite{Catterall:2000rv}. 
In order to avoid the vanishing partition function, we compute the expectation value 
of Hamiltonian \cite{Kanamori:2007yx} at finite temperature (with anti-periodic boundary condition for the fermions) 
and obtain the vacuum energy taking the low-temperature limit.  
Compared with the previous Monte-Carlo results \cite{Kanamori:2007yx}, 
our results 
show good agreement with the known results for a wide range of coupling constants.

The rest of this paper is organized as follows. 
Section \ref{sec:SQM} describes ${\cal N}=2$ supersymmetric quantum mechanics with the path integral formulation
and an introduction to the direct computational method based on the transfer matrix. 
Then, we see the expectation value of Hamiltonian for the cubic superpotential in section \ref{sec:results}.  
A summary and future perspectives are discussed in section \ref{sec:summary}.

\section{SUSY QM on the lattice and the computational method}
\label{sec:SQM}

\subsection{SUSY QM}
\label{sec:H_susy_qm}
%\subsection{Hamiltonian and Euclidean path integral formulation}

${\cal N}=2$ supersymmetric quantum mechanics  is described by the Hamiltonian
\cite{Witten:1981nf,Witten:1982df,Cooper:1994eh},  
\be{
\label{quantum_Hamiltonian}
\hat H = \frac{1}{2} \hat p^2 + \frac{1}{2}W^2(\hat q) - \frac{1}{2}W'(\hat q) \left[\hat b^\dagger, \hat b\right],
}
with the position operator $\hat q$, the momentum operator $\hat p$ 
and the fermionic creation and annihilation operators $ \hat b^\dagger, \hat b$. 
$W(\hat q)$ is any function of $\hat q$, which is called superpotential.

Those operators satisfy the relations
\be{
\left[\hat q,\hat p\right] = i, \m{\ \ \ \ } \left\{\hat b,\hat b^\dagger\right\} = 1, \m{\ \ \ \ } 
\hat b^2 = ({\hat b}^\dag)^2 = 0,
}
and act the state vectors as
%The system is described by the basis $\cket{x,\pm}$ which is characterized as
\be{
\hat q\cket{x} = x\cket{x}, \m{\ \ \ \ } \hat b \cket{-} = 0, \m{\ \ \ \ } \hat b^\dagger\cket{-} = \cket{+}.
}
Since $F \equiv \hat b^\dag \hat b$ is the fermionic number operator, 
%$F=\frac{1-\left[b, b^\dagger\right]}{2}$,
%$F={1-\[b^\dag b\]}/{2}$, 
$\cket{+} $ 
is a fermionic state with $F=1$ while $\cket{-} $ is a bosonic state with $F=0$.

Supercharges $ \hat Q$ and $ \hat Q^\dag$ given by
\begin{eqnarray}
&& \hat Q = \hat b (i \hat p + W(\hat q)), \\
&& \hat Q^\dag = \hat b^\dag (-i \hat p + W( \hat q))
\end{eqnarray}
map a fermionic state to a bosonic state and vice versa.
$\hat Q$ and $\hat Q^\dag$ 
commute with $\hat H$ because $\hat H=\frac{1}{2}\{\hat Q,\hat Q^\dag\}$ and $\hat Q^2=\hat Q^{\dag 2}=0$. 
Any energy eigenstate $\cket{\psi} $ has non-negative energy because $E=\cbra{\psi} \hat H \cket{\psi} = |\hat Q\cket{\psi}|^2 \ge 0$. 
SUSY is broken when there are no normalized energy eigenstates with $E=0$.
%Without solving the eigenvalue equation, 

We can learn whether SUSY is broken or not from the Witten index,
\be{
w=
\mathrm{Tr}
\left\{
(-1)^{F}
\mathrm{e}^{-\beta  \hat H}
\right\},
\label{witten_index}
}
where the trace is taken over all normalized states.
Since the states appear as a pair $\cket{\pm} $ with $E \neq 0$ and 
% $\psi_{+}^{E} = \frac{1}{\sqrt{2E}} Q \psi_{-}^{E}$,%  and  $\psi_{-}^{E} = \frac{1}{\sqrt{2E}} Q^\dag \psi_{+}^{E}$,
cancel with each other in (\ref{witten_index}),  we have 
\begin{eqnarray}
w=n^{E=0}_- - n^{E=0}_+,
\end{eqnarray}
where $n^{E=0}_+$ and  $n^{E=0}_-$  are the number of fermionic and bosonic vacuum states.
The index vanishes if and only if $n^{E=0}_+=n^{E=0}_-=0$ in this model. 
We have $w=0$ when SUSY is broken and vice versa. 
\footnote{ 
In the case of  $|W(\pm \infty)|=\infty $, 
SUSY is broken for $W(\infty)\cdot W(-\infty) >0$ and unbroken for $W(\infty)\cdot W(-\infty) < 0$.
For more comprehensive and detailed discussions are shown in the review \cite{Cooper:1994eh}.}

\begin{figure}[tbp]
\begin{center}
  \includegraphics[width=7cm, angle=0]{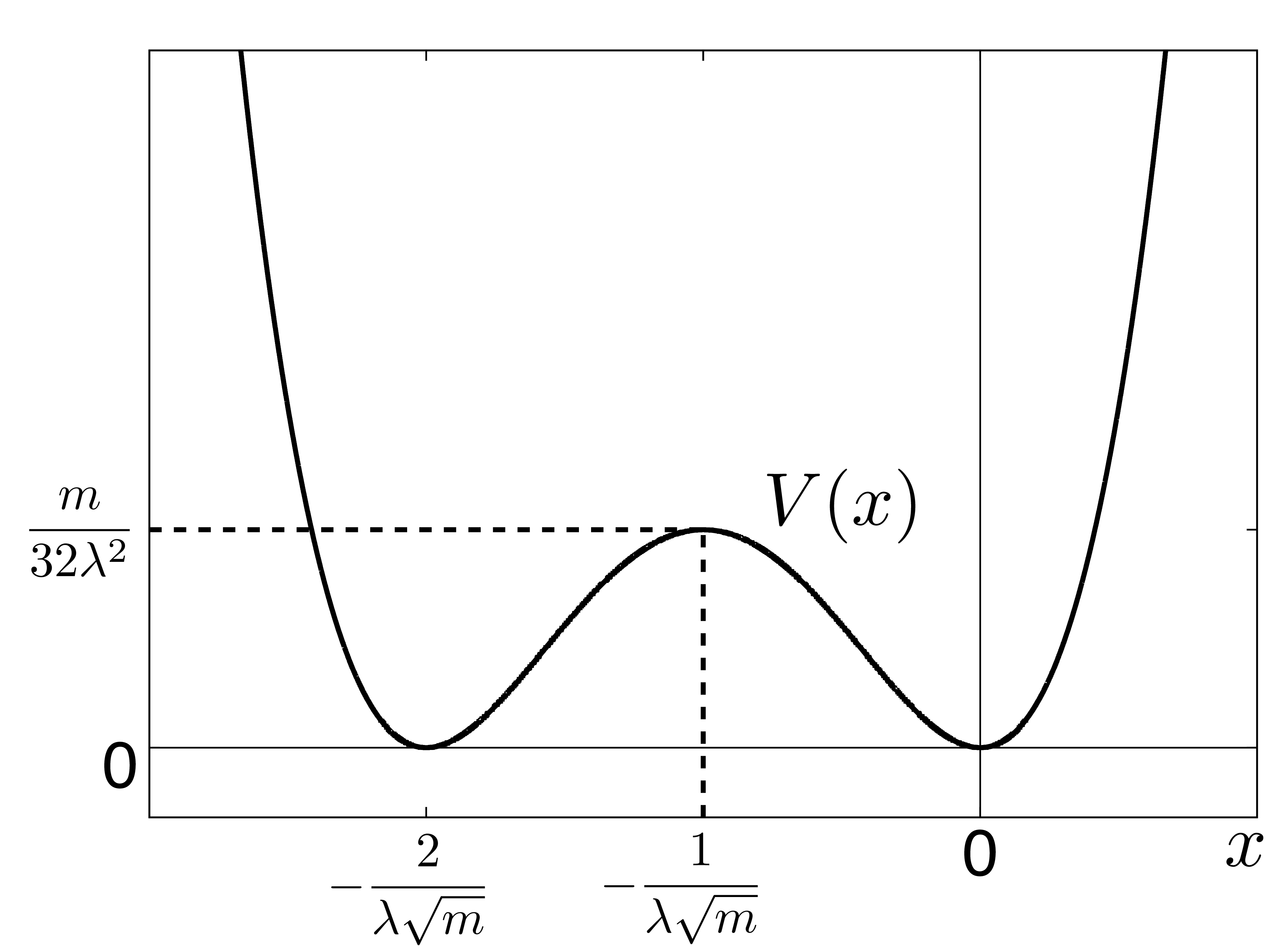}
  \caption{
     $V(x)=\frac{1}{2} W(x)^2$ for a quadratic superpotential.
}
\label{phi4_potential}
\end{center}
\end{figure}

Supersymmetry is broken for a quadratic superpotential \cite{Witten:1981nf,Witten:1982df}, 
\be{
\label{square_potential}
W(x) = m
x
+
g
x^2.
}
Fig.\ref{phi4_potential} shows the potential $V(x)=\frac{1}{2} W(x)^2$ 
with the dimensionless coupling constant $\lambda=g/m^{3/2}$.  
The reason why SUSY is broken is intuitively understood.
The potential has two classical vacua corresponding to two minima 
$x = \{0,-2/(\lambda\sqrt{m})\}$.
These two vacua respect supersymmetry because the classical potential vanishes.
After the quantization, the tunneling between these two vacua 
causes the overlap of the wave function,
 and the true vacuum energy becomes non-zero.
\footnote{The tunneling rate decreases as $\lambda \rightarrow 0$ 
since the potential barrier increases, and the vacuum energy approaches zero.
The vacuum energy can be analytically evaluated by the instanton rate for $\lambda \ll 1$
where the numerical analysis using the importance sampling tends to be difficult.}
%

%This sampling problem is a kind of sign problem, and then we will consider the perturbative regime later in this paper.

The classical action of ${\cal N}=2$ supersymmetric quantum mechanics is given by
\be{
\label{cont_action}
S
=
\integral{0}{\beta}\m{d}t
\left\{
\frac{1}{2}
(\partial_t \phi(t) )^2
+
\frac{1}{2}
W(\phi(t))^2
+
\bar{\psi}(t)
%\overline{\psi}(t)
\big( \partial_t + W'(\phi (t))\big)
\psi (t)
\right\}
}
where $\phi \in \mathbb{R}$ is a bosonic variable and $\psi, \bar \psi$ are fermionic variables 
(one-component Grassmann numbers).
 All variables $\phi,\psi,\bar\psi$  satisfy  the periodic boundary condition such as  
\begin{eqnarray}
\phi(t+\beta) = \phi(t). 
\label{pbc_condt}
\end{eqnarray}
The action is invariant under the supersymmetry transformations,
\begin{eqnarray}
\begin{split}
&
\delta \phi = \epsilon \psi + \bar{\epsilon} \bar{\psi}
\\
& 
\delta \psi = 
-\bar{\epsilon} (\partial_t \phi - W(\phi) )
\\
&
\delta \bar{\psi} = -\epsilon (\partial_t \phi + W(\phi)),
\end{split}
\label{super}
\end{eqnarray}
where $\epsilon$ and $\overline{\epsilon}$ are one-component Grassmann numbers.
The Witten index can be expressed as a path integral form,
\begin{eqnarray}
w = \int  {\rm D} \phi  {\rm D} \bar\psi  {\rm D} \psi \, e^{-S}.
\label{path_integral_w}
\end{eqnarray}
Note again that all field satisfy the periodic boundary condition (\ref{pbc_condt}). 
However, any expectation value is formally ill-defined since the partition function $w=0$ 
for the superpotential (\ref{square_potential}).

To define a well-defined expectation value, 
we consider a statistical system whose partition function is given by
\be{
Z
=
\mathrm{tr}
\left(
\mathrm{e}^{-\beta  \hat H}
\right),
}
with the inverse temperature $\beta=1/T$. 
We also write it down as a path-integral form, 
\begin{eqnarray}
Z = \int  {\rm D}\phi  {\rm D}\bar\psi  {\rm D}\psi \, e^{-S},
\label{path_integral_Z}
\end{eqnarray}
where  $\phi$ satisfy the periodic boundary condition while $\psi(t)$ and $\bar\psi(t)$ satisfy
the anti-periodic boundary condition such as 
\begin{eqnarray}
\psi(t+\beta) = -\psi(t). 
\end{eqnarray}
Note that supersymmetry is explicitly broken due to the anti-periodic boundary condition for the fermions.
The expectation value of an operator ${\cal O}$ is defined in the standard manner as
\begin{eqnarray}
\langle {\cal O} \rangle = \frac{1} {Z}  \int {\rm D}\phi  {\rm D}\bar\psi  {\rm D}\psi \, {\cal O } \, e^{-S},
\label{path_integral_O} 
\end{eqnarray}
since $Z$ is non-zero in general. 
The vacuum energy is obtained as the large $\beta$ limit 
of the expectation value of Hamiltonian $\langle H\rangle$.

\subsection{Lattice theory}
\label{subsec:lattice_theory}

The lattice theory is defined on a lattice such that lattice sites are labeled 
by discretized points with equal intervals, $t\in a\mathbb{Z}$.
The field variables live on the sites and 
lattice boson and fermion are expressed as $\phi_t$ and $\psi_t$, respectively. 
The forward and backward difference operators $\nabla_\pm$ are given by
\begin{eqnarray}
&&\nabla_+ \phi_t = \phi_{t+1}-\phi_t, \\
&& \nabla_- \phi_t = \phi_{t}-\phi_{t-1}.
\end{eqnarray}
Here we set the lattice spacing $a=1$ without loss of generality.

The off-shell expression of the lattice action \cite{Catterall:2000rv} is given by
\begin{eqnarray}
&& 
S_{\rm lat} = \sum_{t=1} ^{N}
\bigg\{
\frac{1}{2}\left(
\nabla_-\phi_t \right)^2
+ 
\frac{1}{2} D_t^2 + iD_t W(\phi_t)
+ \nabla_-\phi_t  W(\phi_t) 
\nonumber \\
&& \hspace{2cm} 
+\overline{\psi}_t \left(
\nabla_- + W'(\phi_t)
\right)\psi_t
\bigg\}.
\label{CG_lattice_action}
\end{eqnarray}
To discuss the SUSY invariance, we first assume that all field satisfy 
the periodic boundary condition such as 
\begin{eqnarray}
\phi_{t+N}= \phi_t,
\label{pbc}
\end{eqnarray}
where $N$ is the size of the lattice.

We consider lattice supersymmetry transformations $\delta=\epsilon Q+  \bar\epsilon \bar Q$: 
\begin{eqnarray}
\begin{split}
& Q\phi = \psi \\
& Q\psi = 0 \\
& Q\bar{\psi} = -\nabla_-\phi-iD\\
& Q D = i\nabla_-\psi,
\end{split}
%\end{eqnarray}
\hspace{1.5cm}
%and 
%\begin{eqnarray}
\begin{split}
&\bar{Q}\phi=\bar{\psi} \\
&\bar{Q}\psi=-\nabla_+\phi +iD \\
&\bar{Q}\bar{\psi}=0 \\
&\bar{Q}D=-i\nabla_+\bar{\psi}.
\end{split}
\end{eqnarray}
These transformations obey $Q^2=\bar Q^2=0$ and $\{Q,\bar Q\}= -(\nabla_+ + \nabla_-)$.

For free theory, 
we can easily show that the lattice action (\ref{CG_lattice_action}) 
is invariant under both of $Q$ and $\bar Q$-transformations.
However, for interacting cases, 
the $Q$-symmetry is only kept on the lattice 
because the action (\ref{CG_lattice_action}) can be written as a $Q$-exact form 
thanks to  the surface term $\nabla_-\phi_t  W(\phi_t)$ while the $\bar Q$-symmetry is explicitly broken.
Numerical and some analytical studies show that 
full supersymmetry is restored in the continuum limit \cite{Catterall:2000rv, Giedt:2004vb, Kadoh:2018ele}.

The Witten index is given by the path integral form such as (\ref{path_integral_w}) 
with the lattice action (\ref{CG_lattice_action}).
The expectation value is, however, ill-defined due to an issue of $0/0$ 
since $w$ vanishes for (\ref{square_potential}) on the lattice.  
So we impose the anti-periodic boundary condition on the fermions as
\begin{eqnarray}
\psi_{t+N}= - \psi_t.
\label{apbc}
\end{eqnarray}
As with (\ref{path_integral_Z}), 
the non-zero partition function is given by 
\begin{eqnarray}
Z = \int {\rm D} \phi {\rm D} \bar \psi {\rm D} \psi {\rm D}D\, \, e^{-S_{\rm lat}},
\label{path_integral_Z_lat}
\end{eqnarray}
with well-defined path integral measures, 
\be{
\label{boson_measure}
\int {\rm D}\phi \equiv \prod_{t=1}^N \int_{-\infty}^{\infty} \frac{d\phi_t}{\sqrt{2\pi}},
}
\be{
\label{fermion_measure}
\int {\rm D}\bar{\psi} {\rm D}\psi \equiv \prod_{t=1}^N \int d\bar{\psi}_t d{\psi}_t,
}
where $d \psi$ and $d\bar\psi$ are the standard Grassmann integral measures  
and the measure for $D$ is given as well as (\ref{boson_measure}).
The expectation value is defined in the same manner as  (\ref{path_integral_O}).

The vacuum energy is obtained from the zero temperature limit of 
$\langle H\rangle$ 
once an appropriate discretization of Hamiltonian is defined.
Since the lattice discretization breaks the translational symmetry, 
we have to deal with the lattice definition of Hamiltonian carefully. 
With the exact $Q$-symmetry, 
the authors of Ref.\cite{Kanamori:2007yx} proposed
a lattice Hamiltonian,
\be
{
H = \frac{1}{2} Q\bar{J},
\label{lattice_hamiltonian}
}
with a supercurrent $\bar{J}$ corresponding to $\bar Q$-transformation, 
\be{
\bar{J} = (\nabla_-\phi - W(\phi))\bar{\psi}.
} 
The $Q$-exactness of (\ref{lattice_hamiltonian}) suggests us 
that the vacuum energy 
with the correct zero point energy is obtained from $\langle H\rangle$.

The explicit form of $H$ is 
\begin{eqnarray}
&& H =-\frac{1}{2} (\nabla_- \phi)^2 -\frac{1}{2} \bar\psi (\nabla_- -W^\prime(\phi)) \psi 
 \nonumber \\
&& \hspace{1cm} 
+ \frac{i}{2} D(W-\nabla_- \phi) + \frac{1}{2} W \nabla_- \phi.
\end{eqnarray}
Using the following identities, 
\begin{eqnarray}
&& \langle \bar\psi_t (\nabla_- + W^\prime(\phi_t))\psi_t \rangle =-1,\\
&& \langle (D+iW) (W-\nabla_- \phi) \rangle =0,
\end{eqnarray}
we obtain
\begin{eqnarray}
\label{explicit_form_H_exp}
\langle H \rangle  = \frac{1}{2} 
\langle -(\nabla_- \phi_t)^2 + W^2(\phi_t) + 2W^\prime(\phi_t) \bar\psi_t \psi_t \rangle + \frac{1}{2}. 
\end{eqnarray}
In the actual computations, we use (\ref{explicit_form_H_exp}) 
to evaluate the expectation value of the lattice Hamiltonian (\ref{lattice_hamiltonian}).

\subsection{The computational method}

We employ the direct computational method based on the transfer matrix, which is 
proposed in Ref.\cite{Kadoh:2018ele},
to evaluate the vacuum energy from the lattice action (\ref{CG_lattice_action}) 
and a lattice definition of Hamiltonian (\ref{lattice_hamiltonian}).

To define the finite dimensional transfer matrix, 
the path integral  is discretized by the Gauss-Hermite quadrature formula: 
\be{
\label{GH_quad}
\int_{-\infty}^{\infty} {d x}
\approx
\sum_{x \in S_K}g_K(x),
}
where  $S_K$ is a set of zero points of $K$-th Hermite polynomials $H_K(x)$ and 
\begin{eqnarray}
g_K(x)= w_K(x) e^{x^2},
\end{eqnarray}
with the weight function $w_K(x)=\frac{2^{K-1} K! \sqrt{\pi}}{K^2 H^2_{K-1}(x)}$. 
We can easily confirm that the standard form of Gauss-Hermite quadrature is obtained 
taking a function $h(x)=e^{-x^2} f(x)$  as the integrand.

Replacing each path integral measure of (\ref{boson_measure}) by (\ref{GH_quad}),
we have an approximation of the partition function (\ref{path_integral_Z_lat}),
\begin{eqnarray}
&& Z \approx  
\sum_{\phi_1 \in S_K} g_K(\phi_1) \ldots \sum_{\phi_N \in S_K}  g_K(\phi_N) 
 \times \left(\frac{1}{\sqrt{2\pi}}\right)^N\, 
\left\{ \prod_{t=1}^N e^{-{\cal L_B}(\phi_t,\phi_{t-1}) }  \right\}
 \nonumber \\
 &&  \hspace{2cm} \times \left\{ 1 + \prod_{t=1}^N(1+W^\prime (\phi_t)) \right\}.
\end{eqnarray}
where 
\be{
{\cal L_B}(x,y)
=
\frac{1}{2}(x - y
+
W(x))^2.
}
Here $D_t$, $\psi_t$ and $\bar\psi_t$ are analytically integrated.

Thus we obtain 
\be{
Z
\approx
\mathrm{Tr}(T_-^N)
+
\mathrm{Tr}(T_+^N),
\label{Z_T_rep}
}
where
\begin{eqnarray}
&&(T_-)_{xy}
=
(1 + W'(x)) (T_+)_{xy},
\\
&&(T_+)_{xy}
=
\sqrt{\frac{g_K(x)g_K(y)}{2\pi}} \,e^{-{\cal L_B}(x,y)}.
\end{eqnarray}
Similarly, the Witten index is also approximately given by
\be{
w
\approx
\mathrm{Tr}(T_-^N)
-
\mathrm{Tr}(T_+^N).
\label{w_T_rep}
}
Note that $T_\pm$ are $K \times K$ matrices which can be regarded as the matrix representations of
 the transfer matrices as discussed in the end of this section. 
Thus we can evaluate the approximate values of $Z$ and $w$ 
from the matrix products of $T_\pm$ in (\ref{Z_T_rep}) and (\ref{w_T_rep}) without using any stochastic process.

With the transfer matrices $T_\pm$, 
(\ref{explicit_form_H_exp}) is also expressed as matrix products, 
\be{
\langle
H
\rangle
=
\frac{1}{Z} \, 
\mathrm{Tr} \bigg\{
\left(H_\mathrm{B}\cdot T_- - {\widetilde T}_+ \right)
T_-^{N-1}
+
\left( H_\mathrm{B} \cdot T_+ \right)
T_+^{N-1}
\bigg\}
+\frac{1}{2}
\label{H_T_rep}
}
where 
\begin{eqnarray}
&&(H_\mathrm{B})_{xy}
=
- \frac{1}{2}(x- y)^2
+ \frac{1}{2} W(x)^2,\\
&& (\widetilde{T}_+)_{xy} = W^\prime (x) ({T}_+)_{xy}, 
\end{eqnarray}
and 
\begin{eqnarray}
(A\cdot B )_{xy}
\equiv
A_{xy} B_{xy}.
\end{eqnarray}
Except for two "dot" products in (\ref{H_T_rep}),
the others are the normal matrix product.

%Note that we can also extract the expectation value from numerical derivative 
%$\langle H\rangle=-\frac{\partial}{\partial \beta}\log{Z_\m{AP}}$ with the inverse temperature $\beta = aN$.
%This definition is an independent way to calculate the expectation value. However, it contains another dependence of lattice spacings $a$ from an approximated numerical derivative of inverse temperature $\beta$.
%Although we check this procedure also reproduce the same result at the continuum limit,
%we use the definition in equation (\ref{Hdef}) for comparison to the previous works.

As discussed in Ref.\cite{Kadoh:2018ele}, 
the rescaling of $\phi$ provides optimized tensors. 
With the rescaling parameter $s$,  
we first change the variable as
\begin{eqnarray}
\Phi_t =s \phi_t,
\end{eqnarray}
before the discretization of the path integral measure.
Repeating the same procedures above,
we find that the tensors are modified as 
\begin{eqnarray}
&&(T_-)_{xy}
=
(1 + W'(x/s)) (T_+)_{xy},
\\
&&(T_+)_{xy}
=
\sqrt{\frac{g_K(x)g_K(y)}{2\pi s^2}} \,e^{-{\cal L_B}(x/s,y/s)}.
\end{eqnarray}
The expectation value of Hamiltonian (\ref{H_T_rep}) is also given 
with the modified $H_\mathrm{B}$ and $\widetilde{T}_+$:   
 \begin{eqnarray}
&&(H_\mathrm{B})_{xy}
=
- \frac{1}{2s^2 }(x- y)^2
+ \frac{1}{2} W(x/s)^2,\\
&& (\widetilde{T}_+)_{xy} = W^\prime (x/s) ({T}_+)_{xy}.
\end{eqnarray}
The accuracy of the approximation is controlled by $K$ and $s$. 
It is expected that the discretized expressions such as (\ref{Z_T_rep}) and (\ref{H_T_rep}) 
approach the correct values as $K$ increases for fixed $s$. 
As explained in the next section, 
a sufficiently large value of $K$ and an optimized value of $s$ are
 chosen to obtain the results as accurate as possible in actual computations.

We finally make some remarks on the relation between $T_{\pm}$ 
and the quantum Hamiltonian $ \hat H$ given by (\ref{quantum_Hamiltonian}).  
Since $\cket{\pm}$ given in section \ref{sec:H_susy_qm}  has two components, 
the wave function $\psi(x)$ is expressed as two column vector,
\be{
\psi=\left(
\begin{array} {l}
\psi_+  \\
\psi_- 
\end{array}
\right).
}
In this representation, the operators $ \hat b, \hat b^\dag, F= \hat b^\dag  \hat b$ 
and the Hamiltonian are also 
expressed as matrix forms,  
\begin{eqnarray}
&&  \hat b =
\begin{pmatrix}
0 & 0 \\
1 & 0
\end{pmatrix}, 
\quad 
 \hat b^\dag =
\begin{pmatrix}
0 & 1 \\
0 & 0
\end{pmatrix},
\quad
F =
\begin{pmatrix}
1 & 0 \\
0 & 0
\end{pmatrix}
,
\end{eqnarray}
and 
\begin{eqnarray}
 \hat H = 
\begin{pmatrix}
 \hat H_+ & 0 \\
0 &  \hat H_-
\end{pmatrix}
,
%&&Q_= \\
%&& Q^\dag=
\end{eqnarray}
where 
\begin{eqnarray}
% \hat H_\pm= \frac{1}{2}  \hat p^2 + \frac{1}{2} W^2( \hat q) \pm  i W^\prime ( \hat q).  
\hat H_\pm= \frac{1}{2}  \hat p^2 + \frac{1}{2} W^2( \hat q) \mp \frac{1}{2} W^\prime ( \hat q).  
\end{eqnarray}
The  Witten index is then given by 
\begin{eqnarray}
w= \mathrm{Tr}
\left(
\mathrm{e}^{-\beta  \hat H_-} 
\right)
-
\mathrm{Tr}
\left(
\mathrm{e}^{-\beta  \hat H_+} 
\right),
\end{eqnarray}
and the partition function with anti-periodic boundary condition for the fermions is 
\begin{eqnarray}
Z= \mathrm{Tr}
\left(
\mathrm{e}^{-\beta  \hat H_-} 
\right)
+
\mathrm{Tr}
\left(
\mathrm{e}^{-\beta  \hat H_+} 
\right).
\end{eqnarray}
Comparing them with (\ref{w_T_rep}) and (\ref{Z_T_rep}), 
we find that $T_\pm$ can be regarded 
as the transfer matrices associated with two Hamiltonians $ \hat H_\pm$ 
which are ones of the bosonic $(-)$ and fermionic states $(+)$.

\section{Results}
\label{sec:results}

The vacuum energy is evaluated using the lattice model (\ref{CG_lattice_action})
in the case of $\phi^4$ interaction for which SUSY is broken.
We employ the direct numerical method 
based on the transfer matrix representations (\ref{Z_T_rep}) and  (\ref{w_T_rep}).
The expectation value of a lattice Hamiltonian, which is given by (\ref{lattice_hamiltonian}), 
is also evaluated from (\ref{H_T_rep}) for several coupling constants.
We will compare the strong coupling results with the previous Monte-Carlo results  
and the weak coupling results with the analytical solution estimated as the instanton effect.

The $\phi^4$ theory is described by the superpotential (\ref{square_potential}).
Since $g$ has the mass dimension $3/2$,  
$\lambda=g/m^{3/2}$ is the dimensionless coupling.
The lattice spacing is measured in the unit of the physical scale $m$ 
and the continuum limit corresponds to $am \rightarrow 0$. 
The physical lattice size is given by $\beta m (= N ma)$ where $N$ is the number of lattice sites.
The  computational parameters employed in this paper are summarized in Table \ref{para_table}. 
\begin{table}
\begin{center}
\begin{tabular}{cc} 
\hline\hline
\multicolumn{2}{c} {
$10 \le \lambda$, $\beta m = 6.4$}  \\
\hline
$am$ & $N$  \\
\hline
0.0040&
1600\\
0.0036&
1778\\
0.0032&
2000\\
0.0028&
2286\\
0.0024&
2667\\
0.0020&
3200\\
0.0016&
4000\\
0.0012&
5333\\
0.0008&
8000\\
0.0004&
16000\\
\hline\hline
\end{tabular}
\hspace{1cm}
%\end{center}
%\begin{center}
\begin{tabular}{cc} 
\hline\hline
\multicolumn{2}{c} {
$\lambda < 10$, $\beta m = 64$}  \\
\hline\hline
$am$ & $N$  \\
\hline
0.050&
1280\\
0.045&
1422\\
0.040&
1600\\
0.035&
1828\\
0.030&
2134\\
0.025&
2560\\
0.020&
3200\\
\hline\hline
\end{tabular}
\end{center}
\caption{
  Parameters used in the computations of $\phi^4$ theory. 
  %We use $\beta m = 6.4$ for $\lambda \ge 10$ and . 
  The lattice spacing $am$ is uniquely given for each $N$ as $am=\beta m/N$,  
  ({\it e.g.}) $am = 0.00359955...$ for $N = 1778$ is shown as $0.0036$ in the table. 
For all listed parameters, we use $K_\m{GH}=200$. For all $am$, 
we set the rescaling parameter $s=0.22$ for $10\le\lambda$ and $s=0.3$ 
for $0.3<\lambda<10$. 
For $\lambda \le 0.3$, we also use $s=0.3$ for $0.035<am$ 
and $s = 0.22$ for $am \le 0.035$.
}
\label{para_table}
\end{table}

$K_\m{GH}$ which is the order of 
Gauss-Hermite quadrature and the rescaling parameter $s$ should be tuned
to reduce the systematic error from the numerical integration 
as well as the case of $\phi^6$ theory studied in \cite{Kadoh:2018ele}.

\begin{figure}[tbp]
\begin{center}
  \includegraphics[width=7cm, angle=-90]{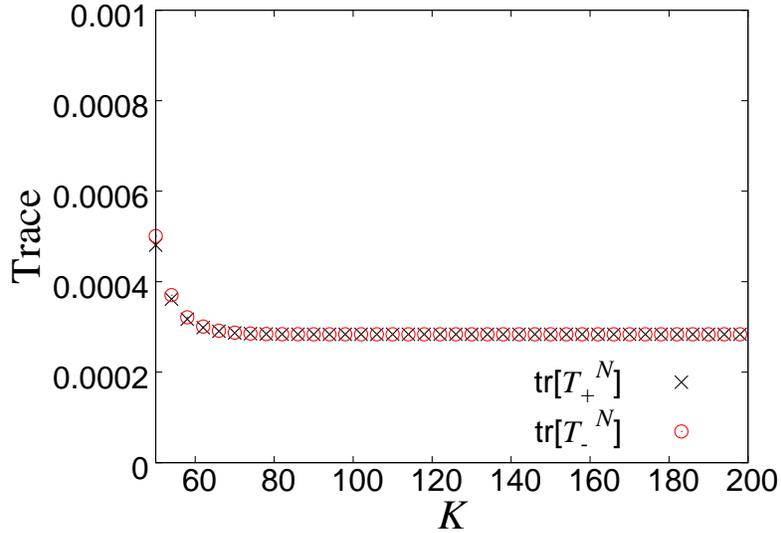}
  \caption{
    Trace of $T_+^N$ (cross) and $T_-^N$ (circle) for  $50\leq K_\m{GH}\leq200$ and $s=0.22$
with a fixed parameters set,  $N=3200$,  $am = 0.002$, $\lambda = 10$, in $\phi^4$ theory.
}
\label{trace_n2}
\end{center}
\end{figure}
Figure \ref{trace_n2} shows the trace of $T_\pm^N$ for
 $50\leq K_\m{GH}\leq200$ and $s=0.22$
with fixed $N=3200$, 
$am = 0.002$, $\lambda = 10$
as an example.
The approximation is improved as $K_\m{GH}$ increases for fixed $s$. 
As seen in the figure, $\m{tr}(T_+^N)$ and $\m{tr}(T_-^N)$ converge to the same value.
This suggests us that    
the Witten index $w=\m{tr}(T_-^N)-\m{tr}(T_+^N)$ vanishes. 
For all parameters in this paper,
we find that $K_\m{GH} = 200$ is large enough to obtain the converged results from the similar studies of Figure \ref{trace_n2}.
We also find that $s=0.22$ realizes $\m{tr}(T_+^N)/\m{tr}(T_-^N) = 1$ 
within the precision $10^{-9}$ for the present parameters set.

\begin{figure}[tbp]
\begin{center}
  \includegraphics[width=7cm, angle=-90]{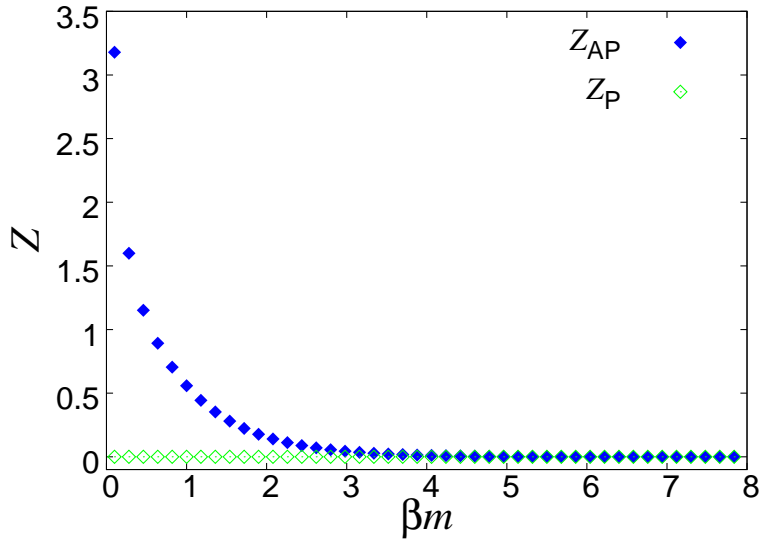}
  \caption{
    Witten index $w=Z_\m{P}$ (open square)  and the partition function 
with the anti-periodic boundary condition on the fermions 
$Z=Z_\m{AP}$ (filled square) for several $\beta m$ in $\phi^4$ theory 
at $am = 0.002$, $\lambda = 10$ with fixed $K_\m{GH}=200$ and $s=0.22$.
}
\label{Z_n2}
\end{center}
\end{figure}

Figure \ref{Z_n2} shows the Witten index and the partition function 
with the anti-periodic boundary condition 
for the fermions  $Z=\m{tr}(T_+^N)+\m{tr}(T_-^N)$ for several lattice size $\beta m$ 
with fixing $K_\m{GH} = 200$ and $s=0.22$. 
As seen in this figure, the Witten index vanishes for all $\beta$ within small errors, 
which are actually of $O(10^{-15})$.

%Although $w$ is not exactly zero due to the systematic error, it can be regarded as zero
%from the numerical integration and the limit of machine precision, the result can be understood as the $Z_\m{P} = 0\pm O(10^{-15})$.
%

The partition function is zero if we impose the periodic boundary condition for the fermions 
since it is the Witten index that vanishes for $\phi^4$ theory. 
To avoid an issue of ill-defined expectation value such as $0/0$,  
we evaluate the expectation value of Hamiltonian $\langle H\rangle$ 
imposing the anti-periodic boundary condition for the fermions with $\beta=1/T$ 
as explained in section \ref{subsec:lattice_theory}. 
The vacuum energy $\epsilon_0$ is evaluated by taking the zero temperature and continuum limit as
\begin{eqnarray}
\epsilon_0 = \lim_{a \rightarrow 0} \lim_{\beta \rightarrow \infty}  \langle H\rangle.
\end{eqnarray}
We use a fixed large value of $\beta m$ 
and a continuum extrapolation with the polynomial functions  
to obtain $\epsilon_0$  in the following.

%\be{
%\frac{\langle H\rangle}{m}
%=
%-\frac{1}{am}\frac{\partial}{\partial N}\log{Z_\m{AP}}.
%}
%

\begin{figure}[tbp]
\begin{center}
  \includegraphics[width=7cm, angle=-90]{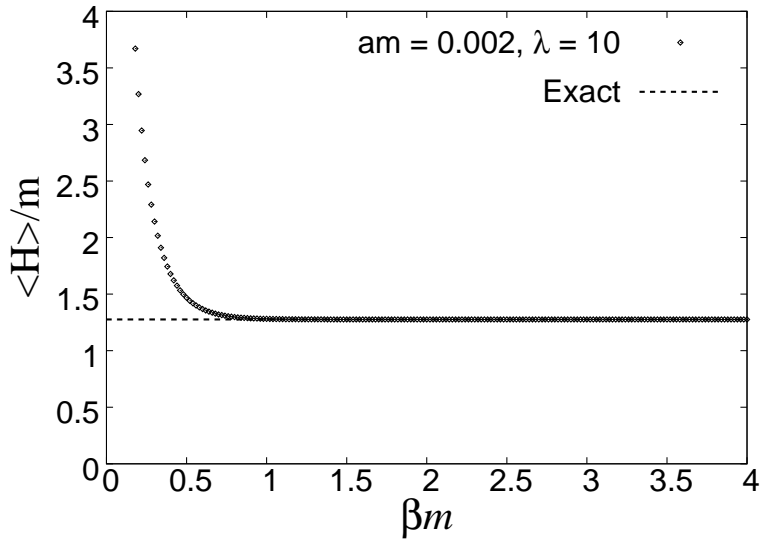}
  \caption{
    Expectation value of the Hamiltonian against $\beta m$ for a strong coupling $\lambda = 10$ 
    with fixed $am = 0.002$. 
    The anti-periodic boundary condition is imposed for the fermions.
    We use $K_\m{GH}=200$ and $s=0.22$ to tune the computational method.
}
\label{hamiltonian}
\end{center}
\end{figure}

Figure \ref{hamiltonian} shows $\langle H\rangle/m$ 
for a strong coupling $\lambda =10$ 
and $0.1\leq \beta m\leq 4$ at $am = 0.002$.
The numerical values of $\langle H\rangle/m$ converge 
as temperature decreases, $\beta \rightarrow \infty$.
We then find that 
$\beta m = 6.4$ is large enough to obtain the results 
within negligible finite temperature effects for  $\lambda =10$. 
%
%
%Although this result is not taken the continuum limit, 
%the result at large lattice size $\langle H(N=1023)\rangle/m = 1.27691$ 
%is crose to exact result $\langle H_\m{exact}\rangle/m = 1.27616$.
%There are already small slope to continuum limit, {\it e.g.} $\langle H(am = 0.02)\rangle/m = 1.27919$ 
%and $\langle H(am = 0.006)\rangle/m =  1.27645$ for larger and smaller lattice spacings.
%
%
With fixed $\beta m = 6.4$, we compute $\langle H\rangle/m$ 
for several $am$ to take the continuum limit.

\begin{figure}[tbp]
\begin{center}
  \includegraphics[width=8cm, angle=-90]{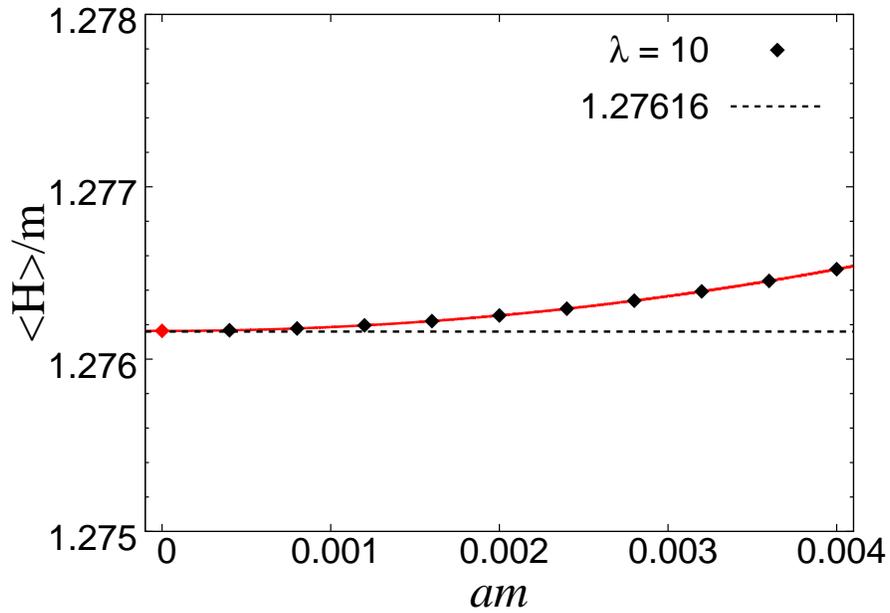}
  \caption{
    Continuum limit of  $\langle H\rangle/m$ for $\lambda = 10$ in $\phi^4$ theory.
We fix the lattice size $\beta m = 6.4$. The anti-periodic boundary 
condition is imposed on the fermions. 
$K_\m{GH}=200$ and $s=0.22 $ are used as the parameters 
associated with the computational method.
}
\label{Hextrap_n2}
\end{center}
\end{figure}

In Figure \ref{Hextrap_n2},
the continuum extrapolation of $\langle H\rangle/m$ with fixed $\beta=6.4$ is shown. 
We can smoothly extrapolate $\langle H\rangle_{\beta=6.4}/m$ 
to the  continuum limit fitting ten finest points with a fit function $f(am)=c_0 + c_1 (am) + c_2 (am)^2$.
We estimate a systematic error from the difference among 
the fit results for five and ten finest points with and without one more higher order term $c_3 (am)^3$.
The maximum difference is used as the systematic error of fittings. 

Table \ref{fit_result_strong} shows the result of the vacuum energy for $\lambda=10$. 
The numerical value is known as $\epsilon_0/m=1.27616$ 
which is obtained by a numerical diagonalization of Hamiltonian~\cite{Balsa:1984eg, Kanamori:2007yx}.
Our extrapolated value  $\langle H\rangle/m = 1.2761637(1)$ 
is consistent with the known result.  
In Ref.\cite{Kanamori:2007yx}, the Monte-Carlo result
with the linear extrapolation function does not reproduce $\epsilon_0/m=1.27616$.
Since the coefficient $c_1$ is consistent with zero, 
the linear extrapolation of \cite{Kanamori:2007yx} may suffer from significant systematic errors. 

\begin{figure}[tbp]
\begin{center}
  \includegraphics[width=7cm, angle=-90]{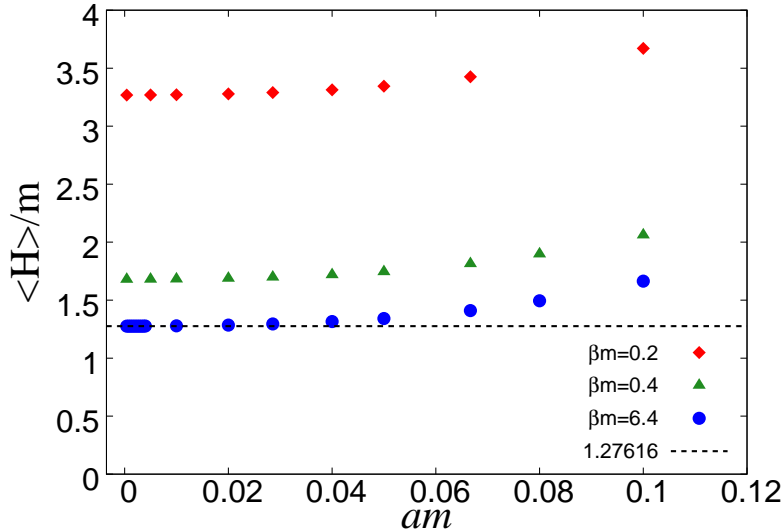}
  \caption{
    Expectation value of Hamiltonian for $\lambda=10$ in $\phi^4$ theory.
    The anti-periodic boundary condition is imposed on the fermions. 
    We use $K_\m{GH}=200$ and $s=0.22$.
    }
\label{Hextrap_n2_2}
\end{center}
\end{figure}

In Figure \ref{Hextrap_n2_2}, we also show
the same scale plot as presented in Ref.\cite{Kanamori:2007yx}. 
Our results and the Monte-Carlo 
ones are consistent with each other. 
The origin of $10$\% discrepancy  
does not come from the definitions of lattice action (\ref{CG_lattice_action}) 
and lattice Hamiltonian (\ref{lattice_hamiltonian}) 
but from systematic errors associated with linear extrapolations as the authors of Ref.\cite{Kanamori:2007yx} discussed. 
We can conclude that the lattice Hamiltonian proposed in Ref.\cite{Kanamori:2007yx} 
works properly.

\begin{table}
\begin{center}
\begin{tabular}{cccc} 
\hline\hline
&$c_0$ &$c_1$ &$c_2$  \\
$\langle H\rangle/m$
&1.2761637(1) 
&-0.00010(14)   
&22.41(8)   \\
\hline\hline
\end{tabular}
\end{center}
\caption{
  Fit result of continuum extrapolation for $\lambda=10$ in $\phi^4$ theory with fixed $\beta m=6.4$.  
The lowest ten data points of $\langle H\rangle/m$ are used for the fit with a fit function 
$f(am) = c_0 + c_1 (ma) +c_2 (ma)^2$. The coefficient $c_0$ gives the vacuum energy $\epsilon_0/m$.
}
\label{fit_result_strong}
\end{table}

We evaluate the vacuum energy for weak couplings for 
which the numerical method becomes more severe as the potential barrier of the double well glows.
The vacuum energy can be analytical evaluated through 
the analysis of the instanton rate \cite{Salomonson:1981ug}:
\be{
\frac{\epsilon_{0,{\rm inst}}}{m}
=
\frac{1}{2\pi}
\mathrm{e}^{-\frac{1}{3\lambda^2}}
(1 + c_1\lambda^2) \qquad  (\lambda \ll 1)
\label{instanton_result}
}
We use a ratio of $\epsilon_0$ to (\ref{instanton_result}) at the leading order,
\begin{eqnarray}
R= \frac{2\pi \epsilon_0}{m} \mathrm{e}^{\frac{1}{3\lambda^2}},
\end{eqnarray}
to test whether our computations reproduce the instanton effect (\ref{instanton_result}) or not.
$R$ should behave as $R =1+d_1 \lambda^2+\ldots$ for $\lambda \ll 1$.

\begin{figure}[tbp]
\begin{center}
  \includegraphics[width=7cm, angle=-90]{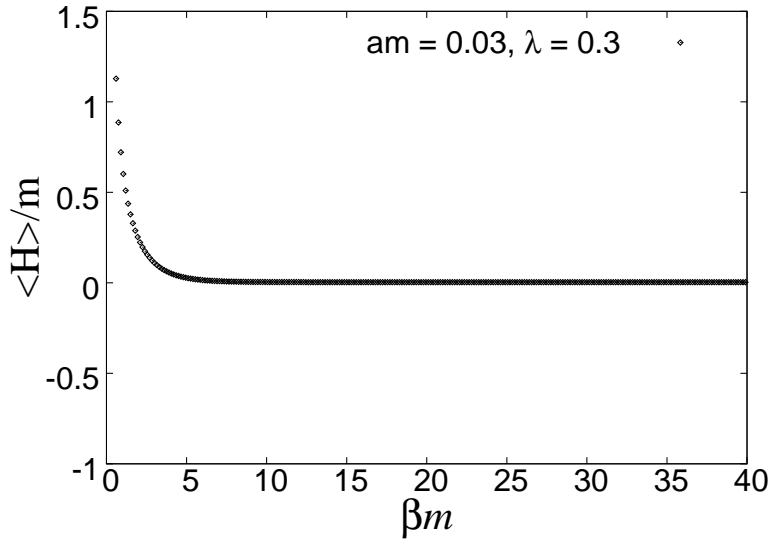}
  \caption{
    Expectation value of the Hamiltonian against $\beta m$ for a weak coupling $\lambda = 0.3$ 
    with fixed $am = 0.03$. 
    The anti-periodic boundary condition is imposed for the fermions.
    We use $K_\m{GH}=200$ and $s=0.3$ to tune the computational method.
}
\label{hamiltonian_weak}
\end{center}
\end{figure}

Figure \ref{hamiltonian_weak} shows the $\langle H\rangle/m$ 
for a weak coupling $\lambda =0.3$ 
and $0.15\leq \beta m \leq 40$ at $am = 0.03$.
The numerical values of $\langle H\rangle/m$ converge 
as temperature decreases as well as the case of $\lambda=10$. 
We take $\beta m = 64$ as  a sufficiently large value of the physical volume for the weak coupling,
which is larger than that for $\lambda=10$. 
With fixed $\beta m = 64$, we compute $\langle H\rangle/m$ 
for different lattice spacings $am$ to take the continuum limit.

\begin{figure}[tbp]
\begin{center}
  \includegraphics[width=7cm, angle=-90]{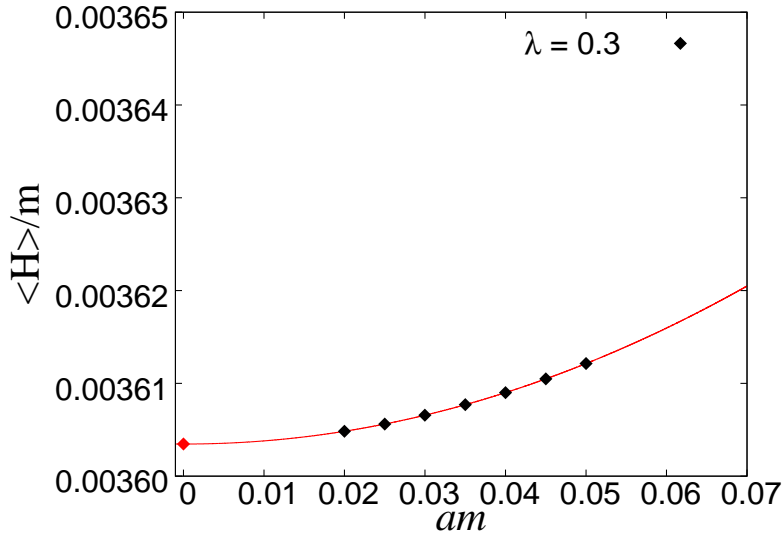}
  \caption{
   Continuum limit of  $\langle H\rangle/m$ for $\lambda = 0.3$ in $\phi^4$ theory. 
   $\beta m = 64$ is fixed.
The fermions satisfy the anti-periodic boundary condition.  
$K_\m{GH}=200$ and $s=0.3$ are used.
    }
\label{each_extrap}
\end{center}
\end{figure}

Figure \ref{each_extrap} shows the extrapolation of $\langle H\rangle/m$ 
to the continuum limit for $\beta m=64$ and $\lambda = 0.3$.
\begin{table}
\begin{center}
\begin{tabular}{cccc} 
\hline\hline
&$c_0$ &$c_1$ &$c_2$  \\
$\langle H\rangle/m$
&0.00360346(2)
&0.0000011(17)
&0.00349(5)  \\
\hline\hline
\end{tabular}
\end{center}
\caption{
  Fit result of continuum extrapolation for $\lambda=0.3$ in $\phi^4$ theory with fixed $\beta m=64$. 
  We use the quadratic function $f(am) = c_0 + c_1 (ma) +c_2 (ma)^2$ to fit $\langle H\rangle/m$
  where $c_0$ gives the vacuum energy $\epsilon_0/m$.
}
\label{fit_result_weak}
\end{table}
Table \ref{fit_result_weak} shows the fit result 
of the continuum extrapolation with the quadratic function.   
Compared with Table \ref{fit_result_strong}, we again confirm that $c_1$ is zero within the error.
Although the lattice spacings are larger than those of $\lambda=10$, 
the extrapolation effectively works within the precision of approximately $0.01$\%. 
This is because $c_2/c_0 \simeq 1$ for $\lambda=0.3$ although $c_2/c_0 \simeq 18$ for $\lambda=10$.
The systematic error originated from the continuum limit 
is actually much smaller than that from the $\lambda \tendto 0$ limit.

\begin{figure}[tbp]
\begin{center}
  \includegraphics[width=7cm, angle=-90]{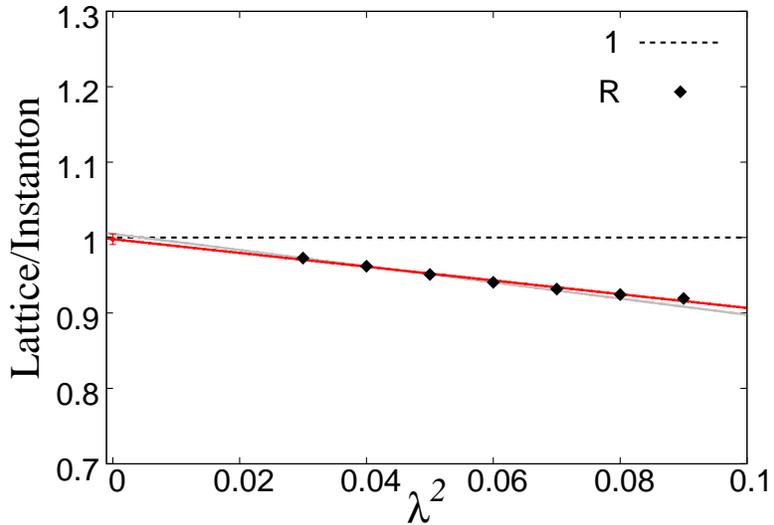}
  \caption{
    $R$ against $\lambda^2$ for fixed $\beta m = 64$.
    }
\label{instanton}
\end{center}
\end{figure}

In Figure \ref{instanton},  we take $\lambda \tendto 0$ limit 
using the linear function $g(\lambda^2) = d_0 + d_1 \lambda^2$.
The error is estimated by the difference from the fit for the finest four data points. 
We find that $R = 0.998(7) - 0.91(16)\lambda^2$. 
The instanton effect is thus precisely reproduced 
from the lattice model (\ref{CG_lattice_action}) with
the $Q$-exact definition of lattice Hamiltonian (\ref{lattice_hamiltonian}) using our method.

\begin{figure}[tbp]
\begin{center}
  \includegraphics[width=7cm, angle=-90]{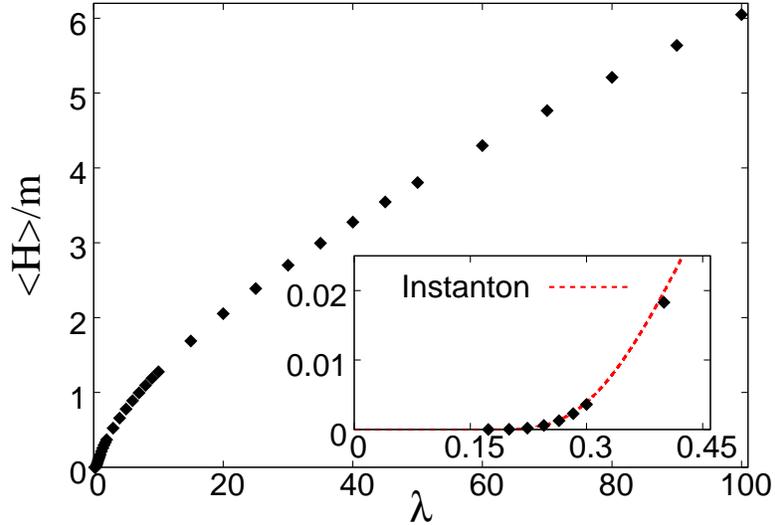}
  \caption{
Vacuum energy estimated from the lattice model (\ref{CG_lattice_action})
and a lattice Hamiltonian (\ref{lattice_hamiltonian}) in $\phi^4$ theory.  
    }
\label{lambda_range}
\end{center}
\end{figure}

Finally, we present the vacuum energy for various coupling constants, 
$\sqrt{0.03}\leq\lambda\leq 100$ in Figure \ref{lambda_range}.
We choose the rescaling parameter $s$ of the computational method and the lattice spacing appropriately 
as shown in Table \ref{para_table}.   
The vacuum energy is obtained in high accuracy as shown 
in Table \ref{fit_result_strong} and Table \ref{fit_result_weak} for $\lambda=10$ and $\lambda=0.3$, respectively.
Although the error becomes larger and larger as $\lambda$ decreases, 
$\epsilon_0$ is obtained  within the error of the order of $0.1$\% 
for the smallest value of coupling constant, $\lambda^2=0.03$.

\section{Summary}
\label{sec:summary}

We have evaluated the vacuum energy in $N=2$ SUSY QM for the double-well potential 
($\phi^4$-theory) using the direct computational method proposed in \cite{Kadoh:2018ele}. 
Since the partition function with the periodic boundary condition for the fermions vanishes, 
we have measured the expectation of Hamiltonian at finite temperature 
and obtain the vacuum energy by taking the low temperature and continuum limit.
The obtained energy coincides with the known results for various coupling constants.
With the studies for the SUSY unbroken case \cite{Kadoh:2018ele}, 
we find that the employed method works properly
with and without the SUSY breaking.

These results also establish the methodology of defining  a lattice Hamiltonian 
in low-dimensional lattice SUSY actions.  
Since the accurate results are provided  for very fine lattice spacings, 
we can precisely study how the lattice artifacts are appeared 
and controlled in the presence of the physical SUSY breaking.
Those kinds of information are very useful in constructing 
higher dimensional SUSY lattice models and 
in studying the non-perturbative mechanism of SUSY breaking.

%\appendix
%\input{app1}

\section*{Acknowledgments}
We would like to thank Hiroto So, Hiroshi Suzuki, Issaku Kanamori and Takeru Kamei 
for their valuable comments.
D.K. is supported by JSPS KAKENHI Grant (No. 16K05328).
K.N. is supported partly by the Grant-in-Aid 
for JSPS (Japan Society for the Promotion of Science) Research Fellow (No. 18J11457).

\bibliographystyle{elsarticle-num}

\bibliography{Refs}

\begin{thebibliography}{10}
\expandafter\ifx\csname url\endcsname\relax
  \def\url#1{\texttt{#1}}\fi
\expandafter\ifx\csname urlprefix\endcsname\relax\def\urlprefix{URL }\fi
\expandafter\ifx\csname href\endcsname\relax
  \def\href#1#2{#2} \def\path#1{#1}\fi

\bibitem{Witten:1981nf}
E.~Witten, {Dynamical Breaking of Supersymmetry}, Nucl. Phys. B188 (1981) 513.
\newblock \href {http://dx.doi.org/10.1016/0550-3213(81)90006-7}
  {\path{doi:10.1016/0550-3213(81)90006-7}}.

\bibitem{Witten:1982df}
E.~Witten, {Constraints on Supersymmetry Breaking}, Nucl. Phys. B202 (1982)
  253.
\newblock \href {http://dx.doi.org/10.1016/0550-3213(82)90071-2}
  {\path{doi:10.1016/0550-3213(82)90071-2}}.

\bibitem{Beccaria:2004pa}
M.~Beccaria, M.~Campostrini, A.~Feo, {Supersymmetry breaking in two-dimensions:
  The Lattice N = 1 Wess-Zumino model}, Phys. Rev. D69 (2004) 095010.
\newblock \href {http://arxiv.org/abs/hep-lat/0402007}
  {\path{arXiv:hep-lat/0402007}}, \href
  {http://dx.doi.org/10.1103/PhysRevD.69.095010}
  {\path{doi:10.1103/PhysRevD.69.095010}}.

\bibitem{Kanamori:2007yx}
I.~Kanamori, F.~Sugino, H.~Suzuki, {Observing dynamical supersymmetry breaking
  with euclidean lattice simulations}, Prog. Theor. Phys. 119 (2008) 797--827.
\newblock \href {http://arxiv.org/abs/0711.2132} {\path{arXiv:0711.2132}},
  \href {http://dx.doi.org/10.1143/PTP.119.797}
  {\path{doi:10.1143/PTP.119.797}}.

\bibitem{Wozar:2011gu}
C.~Wozar, A.~Wipf, {Supersymmetry Breaking in Low Dimensional Models}, Annals
  Phys. 327 (2012) 774--807.
\newblock \href {http://arxiv.org/abs/1107.3324} {\path{arXiv:1107.3324}},
  \href {http://dx.doi.org/10.1016/j.aop.2011.11.015}
  {\path{doi:10.1016/j.aop.2011.11.015}}.

\bibitem{Catterall:2015tta}
S.~Catterall, A.~Veernala, {Spontaneous supersymmetry breaking in two
  dimensional lattice super QCD}, JHEP 10 (2015) 013.
\newblock \href {http://arxiv.org/abs/1505.00467} {\path{arXiv:1505.00467}},
  \href {http://dx.doi.org/10.1007/JHEP10(2015)013}
  {\path{doi:10.1007/JHEP10(2015)013}}.

\bibitem{Catterall:2017xox}
S.~Catterall, R.~G. Jha, A.~Joseph, {Nonperturbative study of dynamical SUSY
  breaking in N=(2,2) Yang-Mills theory}, Phys. Rev. D97~(5) (2018) 054504.
\newblock \href {http://arxiv.org/abs/1801.00012} {\path{arXiv:1801.00012}},
  \href {http://dx.doi.org/10.1103/PhysRevD.97.054504}
  {\path{doi:10.1103/PhysRevD.97.054504}}.

\bibitem{Dondi:1976tx}
P.~H. Dondi, H.~Nicolai, {Lattice Supersymmetry}, Nuovo Cim. A41 (1977) 1.
\newblock \href {http://dx.doi.org/10.1007/BF02730448}
  {\path{doi:10.1007/BF02730448}}.

\bibitem{Elitzur:1982vh}
S.~Elitzur, E.~Rabinovici, A.~Schwimmer, {Supersymmetric Models on the
  Lattice}, Phys. Lett. 119B (1982) 165.
\newblock \href {http://dx.doi.org/10.1016/0370-2693(82)90269-6}
  {\path{doi:10.1016/0370-2693(82)90269-6}}.

\bibitem{Elitzur:1983nj}
S.~Elitzur, A.~Schwimmer, {$N=2$ Two-dimensional {Wess-Zumino} Model on the
  Lattice}, Nucl. Phys. B226 (1983) 109--120.
\newblock \href {http://dx.doi.org/10.1016/0550-3213(83)90465-0}
  {\path{doi:10.1016/0550-3213(83)90465-0}}.

\bibitem{Cecotti:1982ad}
S.~Cecotti, L.~Girardello, {Stochastic Processes in Lattice (Extended)
  Supersymmetry}, Nucl. Phys. B226 (1983) 417--428.
\newblock \href {http://dx.doi.org/10.1016/0550-3213(83)90200-6}
  {\path{doi:10.1016/0550-3213(83)90200-6}}.

\bibitem{Sakai:1983dg}
N.~Sakai, M.~Sakamoto, {Lattice Supersymmetry and the Nicolai Mapping}, Nucl.
  Phys. B229 (1983) 173--188.
\newblock \href {http://dx.doi.org/10.1016/0550-3213(83)90359-0}
  {\path{doi:10.1016/0550-3213(83)90359-0}}.

\bibitem{Golterman:1988ta}
M.~F.~L. Golterman, D.~N. Petcher, {A Local Interactive Lattice Model With
  Supersymmetry}, Nucl. Phys. B319 (1989) 307--341.
\newblock \href {http://dx.doi.org/10.1016/0550-3213(89)90080-1}
  {\path{doi:10.1016/0550-3213(89)90080-1}}.

\bibitem{Catterall:2000rv}
S.~Catterall, E.~Gregory, {A Lattice path integral for supersymmetric quantum
  mechanics}, Phys. Lett. B487 (2000) 349--356.
\newblock \href {http://arxiv.org/abs/hep-lat/0006013}
  {\path{arXiv:hep-lat/0006013}}, \href
  {http://dx.doi.org/10.1016/S0370-2693(00)00835-2}
  {\path{doi:10.1016/S0370-2693(00)00835-2}}.

\bibitem{Kikukawa:2002as}
Y.~Kikukawa, Y.~Nakayama, {Nicolai mapping versus exact chiral symmetry on the
  lattice}, Phys. Rev. D66 (2002) 094508.
\newblock \href {http://arxiv.org/abs/hep-lat/0207013}
  {\path{arXiv:hep-lat/0207013}}, \href
  {http://dx.doi.org/10.1103/PhysRevD.66.094508}
  {\path{doi:10.1103/PhysRevD.66.094508}}.

\bibitem{Giedt:2004qs}
J.~Giedt, E.~Poppitz, {Lattice supersymmetry, superfields and renormalization},
  J. High Energy Phys. 09 (2004) 029.
\newblock \href {http://arxiv.org/abs/hep-th/0407135}
  {\path{arXiv:hep-th/0407135}}, \href
  {http://dx.doi.org/10.1088/1126-6708/2004/09/029}
  {\path{doi:10.1088/1126-6708/2004/09/029}}.

\bibitem{Feo:2004kx}
A.~Feo, {Predictions and recent results in SUSY on the lattice}, Mod. Phys.
  Lett. A19 (2004) 2387--2402.
\newblock \href {http://arxiv.org/abs/hep-lat/0410012}
  {\path{arXiv:hep-lat/0410012}}, \href
  {http://dx.doi.org/10.1142/S0217732304015749}
  {\path{doi:10.1142/S0217732304015749}}.

\bibitem{Kato:2008sp}
M.~Kato, M.~Sakamoto, H.~So, {Taming the Leibniz Rule on the Lattice}, JHEP 05
  (2008) 057.
\newblock \href {http://arxiv.org/abs/0803.3121} {\path{arXiv:0803.3121}},
  \href {http://dx.doi.org/10.1088/1126-6708/2008/05/057}
  {\path{doi:10.1088/1126-6708/2008/05/057}}.

\bibitem{Kadoh:2009sp}
D.~Kadoh, H.~Suzuki, {Supersymmetric nonperturbative formulation of the WZ
  model in lower dimensions}, Phys. Lett. B684 (2010) 167--172.
\newblock \href {http://arxiv.org/abs/0909.3686} {\path{arXiv:0909.3686}},
  \href {http://dx.doi.org/10.1016/j.physletb.2010.01.022}
  {\path{doi:10.1016/j.physletb.2010.01.022}}.

\bibitem{Kadoh:2010ca}
D.~Kadoh, H.~Suzuki, {Supersymmetry restoration in lattice formulations of 2D
  $\mathcal{N}=(2,2)$ WZ model based on the Nicolai map}, Phys. Lett. B696
  (2011) 163--166.
\newblock \href {http://arxiv.org/abs/1011.0788} {\path{arXiv:1011.0788}},
  \href {http://dx.doi.org/10.1016/j.physletb.2010.12.012}
  {\path{doi:10.1016/j.physletb.2010.12.012}}.

\bibitem{Kato:2013sba}
M.~Kato, M.~Sakamoto, H.~So, {A criterion for lattice supersymmetry: cyclic
  Leibniz rule}, JHEP 05 (2013) 089.
\newblock \href {http://arxiv.org/abs/1303.4472} {\path{arXiv:1303.4472}},
  \href {http://dx.doi.org/10.1007/JHEP05(2013)089}
  {\path{doi:10.1007/JHEP05(2013)089}}.

\bibitem{Kadoh:2015zza}
D.~Kadoh, N.~Ukita, {General solution of the cyclic Leibniz rule}, PTEP
  2015~(10) (2015) 103B04.
\newblock \href {http://arxiv.org/abs/1503.06922} {\path{arXiv:1503.06922}},
  \href {http://dx.doi.org/10.1093/ptep/ptv140}
  {\path{doi:10.1093/ptep/ptv140}}.

\bibitem{Asaka:2016cxm}
K.~Asaka, A.~D'Adda, N.~Kawamoto, Y.~Kondo, {Exact lattice supersymmetry at the
  quantum level for $N = 2$ Wess-Zumino models in 1- and 2-dimensions}, Int. J.
  Mod. Phys. A31~(23) (2016) 1650125.
\newblock \href {http://arxiv.org/abs/1607.04371} {\path{arXiv:1607.04371}},
  \href {http://dx.doi.org/10.1142/S0217751X16501256}
  {\path{doi:10.1142/S0217751X16501256}}.

\bibitem{Kato:2016fpg}
M.~Kato, M.~Sakamoto, H.~So, {Non-renormalization theorem in a lattice
  supersymmetric theory and the cyclic Leibniz rule}, PTEP 2017~(4) (2017)
  043B09.
\newblock \href {http://arxiv.org/abs/1609.08793} {\path{arXiv:1609.08793}},
  \href {http://dx.doi.org/10.1093/ptep/ptx045}
  {\path{doi:10.1093/ptep/ptx045}}.

\bibitem{DAdda:2017bzo}
A.~D'Adda, N.~Kawamoto, J.~Saito, {An Alternative Lattice Field Theory
  Formulation Inspired by Lattice Supersymmetry}, J. High Energy Phys. 12
  (2017) 089.
\newblock \href {http://arxiv.org/abs/1706.02615} {\path{arXiv:1706.02615}},
  \href {http://dx.doi.org/10.1007/JHEP12(2017)089}
  {\path{doi:10.1007/JHEP12(2017)089}}.

\bibitem{Kadoh:2018ele}
D.~Kadoh, K.~Nakayama, {Direct computational approach to lattice supersymmetric
  quantum mechanics}\href {http://arxiv.org/abs/1803.07960}
  {\path{arXiv:1803.07960}}.

\bibitem{Bergner:2007pu}
G.~Bergner, T.~Kaestner, S.~Uhlmann, A.~Wipf, {Low-dimensional Supersymmetric
  Lattice Models}, Annals Phys. 323 (2008) 946--988.
\newblock \href {http://arxiv.org/abs/0705.2212} {\path{arXiv:0705.2212}},
  \href {http://dx.doi.org/10.1016/j.aop.2007.06.010}
  {\path{doi:10.1016/j.aop.2007.06.010}}.

\bibitem{Schierenberg:2012pb}
S.~Schierenberg, F.~Bruckmann, {Improved lattice actions for supersymmetric
  quantum mechanics}, Phys. Rev. D89~(1) (2014) 014511.
\newblock \href {http://arxiv.org/abs/1210.5404} {\path{arXiv:1210.5404}},
  \href {http://dx.doi.org/10.1103/PhysRevD.89.014511}
  {\path{doi:10.1103/PhysRevD.89.014511}}.

\bibitem{Beccaria:1998vi}
M.~Beccaria, G.~Curci, E.~D'Ambrosio, {Simulation of supersymmetric models with
  a local Nicolai map}, Phys. Rev. D58 (1998) 065009.
\newblock \href {http://arxiv.org/abs/hep-lat/9804010}
  {\path{arXiv:hep-lat/9804010}}, \href
  {http://dx.doi.org/10.1103/PhysRevD.58.065009}
  {\path{doi:10.1103/PhysRevD.58.065009}}.

\bibitem{Catterall:2001fr}
S.~Catterall, S.~Karamov, {Exact lattice supersymmetry: The Two-dimensional N=2
  Wess-Zumino model}, Phys. Rev. D65 (2002) 094501.
\newblock \href {http://arxiv.org/abs/hep-lat/0108024}
  {\path{arXiv:hep-lat/0108024}}, \href
  {http://dx.doi.org/10.1103/PhysRevD.65.094501}
  {\path{doi:10.1103/PhysRevD.65.094501}}.

\bibitem{Giedt:2004vb}
J.~Giedt, R.~Koniuk, E.~Poppitz, T.~Yavin, {Less naive about supersymmetric
  lattice quantum mechanics}, JHEP 12 (2004) 033.
\newblock \href {http://arxiv.org/abs/hep-lat/0410041}
  {\path{arXiv:hep-lat/0410041}}, \href
  {http://dx.doi.org/10.1088/1126-6708/2004/12/033}
  {\path{doi:10.1088/1126-6708/2004/12/033}}.

\bibitem{Giedt:2005ae}
J.~Giedt, {R-symmetry in the Q-exact (2,2) 2-D lattice Wess-Zumino model},
  Nucl. Phys. B726 (2005) 210--232.
\newblock \href {http://arxiv.org/abs/hep-lat/0507016}
  {\path{arXiv:hep-lat/0507016}}, \href
  {http://dx.doi.org/10.1016/j.nuclphysb.2005.08.004}
  {\path{doi:10.1016/j.nuclphysb.2005.08.004}}.

\bibitem{Kastner:2008zc}
T.~Kastner, G.~Bergner, S.~Uhlmann, A.~Wipf, C.~Wozar, {Two-Dimensional
  Wess-Zumino Models at Intermediate Couplings}, Phys. Rev. D78 (2008) 095001.
\newblock \href {http://arxiv.org/abs/0807.1905} {\path{arXiv:0807.1905}},
  \href {http://dx.doi.org/10.1103/PhysRevD.78.095001}
  {\path{doi:10.1103/PhysRevD.78.095001}}.

\bibitem{Bergner:2009vg}
G.~Bergner, {Complete supersymmetry on the lattice and a No-Go theorem}, JHEP
  01 (2010) 024.
\newblock \href {http://arxiv.org/abs/0909.4791} {\path{arXiv:0909.4791}},
  \href {http://dx.doi.org/10.1007/JHEP01(2010)024}
  {\path{doi:10.1007/JHEP01(2010)024}}.

\bibitem{Kawai:2010yj}
H.~Kawai, Y.~Kikukawa, {A Lattice study of N=2 Landau-Ginzburg model using a
  Nicolai map}, Phys. Rev. D83 (2011) 074502.
\newblock \href {http://arxiv.org/abs/1005.4671} {\path{arXiv:1005.4671}},
  \href {http://dx.doi.org/10.1103/PhysRevD.83.074502}
  {\path{doi:10.1103/PhysRevD.83.074502}}.

\bibitem{Kanamori:2010gw}
I.~Kanamori, {A Method for Measuring the Witten Index Using Lattice
  Simulation}, Nucl. Phys. B841 (2010) 426--447.
\newblock \href {http://arxiv.org/abs/1006.2468} {\path{arXiv:1006.2468}},
  \href {http://dx.doi.org/10.1016/j.nuclphysb.2010.08.010}
  {\path{doi:10.1016/j.nuclphysb.2010.08.010}}.

\bibitem{Kamata:2011fr}
S.~Kamata, H.~Suzuki, {Numerical simulation of the $\mathcal{N}=(2,2)$
  Landau-Ginzburg model}, Nucl. Phys. B854 (2012) 552--574.
\newblock \href {http://arxiv.org/abs/1107.1367} {\path{arXiv:1107.1367}},
  \href {http://dx.doi.org/10.1016/j.nuclphysb.2011.09.007}
  {\path{doi:10.1016/j.nuclphysb.2011.09.007}}.

\bibitem{Steinhauer:2014yaa}
K.~Steinhauer, U.~Wenger, {Spontaneous supersymmetry breaking in the 2D
  $\mathcal N=$1 Wess-Zumino model}, Phys. Rev. Lett. 113~(23) (2014) 231601.
\newblock \href {http://arxiv.org/abs/1410.6665} {\path{arXiv:1410.6665}},
  \href {http://dx.doi.org/10.1103/PhysRevLett.113.231601}
  {\path{doi:10.1103/PhysRevLett.113.231601}}.

\bibitem{Kadoh:2016eju}
D.~Kadoh, {Recent progress in lattice supersymmetry: from lattice gauge theory
  to black holes}, PoS LATTICE2015 (2016) 017.
\newblock \href {http://arxiv.org/abs/1607.01170} {\path{arXiv:1607.01170}},
  \href {http://dx.doi.org/10.22323/1.251.0017}
  {\path{doi:10.22323/1.251.0017}}.

\bibitem{Cooper:1994eh}
F.~Cooper, A.~Khare, U.~Sukhatme, {Supersymmetry and quantum mechanics}, Phys.
  Rept. 251 (1995) 267--385.
\newblock \href {http://arxiv.org/abs/hep-th/9405029}
  {\path{arXiv:hep-th/9405029}}, \href
  {http://dx.doi.org/10.1016/0370-1573(94)00080-M}
  {\path{doi:10.1016/0370-1573(94)00080-M}}.

\bibitem{Balsa:1984eg}
R.~Balsa, M.~Plo, J.~G. Esteve, A.~F. Pacheco, {SIMPLE PROCEDURE TO COMPUTE
  ACCURATE ENERGY LEVELS OF A DOUBLE WELL ANHARMONIC OSCILLATOR}, Phys. Rev.
  D28 (1983) 1945--1948.
\newblock \href {http://dx.doi.org/10.1103/PhysRevD.28.1945}
  {\path{doi:10.1103/PhysRevD.28.1945}}.

\bibitem{Salomonson:1981ug}
P.~Salomonson, J.~W. van Holten, {Fermionic Coordinates and Supersymmetry in
  Quantum Mechanics}, Nucl. Phys. B196 (1982) 509--531.
\newblock \href {http://dx.doi.org/10.1016/0550-3213(82)90505-3}
  {\path{doi:10.1016/0550-3213(82)90505-3}}.

\end{thebibliography}

\end{document}